\documentclass[prl,twocolumn,amsmath,amssymb,superscriptaddress,showpacs]{revtex4}

\usepackage{graphicx}
\usepackage{dcolumn}
\usepackage{bm}
\newcommand{\Mtt}{M_{t\overline{t}}}

\newcommand{\dxs}{d\sigma/dM_{t\overline{t}}}
\newcommand{\gev}{\mathrm{GeV}/c^{2}}

\newcommand{\acc}{\mathcal{A}_{i}}

\newcommand{\invfb}{\mathrm{fb}^{-1}}
\newcommand{\ahat}{\hat{A}}
\newcommand{\met}{E\!\!\!/_T }

\newcommand{\fbpergev}{\mathrm{fb/GeV}/c^{2}}

\newcommand{\ppbar}{p\bar{p}}

\newcommand{\ttbar}{t\bar{t}}
\newcommand{\bbbar}{b\bar{b}}
\newcommand{\ccbar}{c\bar{c}}
\newcommand{\GeVc}{\mbox{$\mathrm{GeV\!/\!c}$}}
\newcommand{\GeVcc}{\mbox{$\mathrm{GeV\!/\!c^2}$}}
\newcommand{\Pt}{P_{\mathrm T}}
\newcommand{\Et}{E_{\mathrm T}}

\begin{document}

\title{First Measurement of the $\lowercase{\ttbar}$ Differential Cross Section
$\lowercase{d\sigma/d}M_{\lowercase{{t\overline{t}}}}$ in
$\lowercase{\ppbar}$ Collisions at $\sqrt{\lowercase{s}}=1.96$ TeV}
\vspace*{2.0cm}

\affiliation{Institute of Physics, Academia Sinica, Taipei, Taiwan
11529, Republic of China} \affiliation{Argonne National Laboratory,
Argonne, Illinois 60439} \affiliation{University of Athens, 157 71
Athens, Greece} \affiliation{Institut de Fisica d'Altes Energies,
Universitat Autonoma de Barcelona, E-08193, Bellaterra (Barcelona),
Spain} \affiliation{Baylor University, Waco, Texas  76798}
\affiliation{Istituto Nazionale di Fisica Nucleare Bologna,
$^x$University of Bologna, I-40127 Bologna, Italy}
\affiliation{Brandeis University, Waltham, Massachusetts 02254}
\affiliation{University of California, Davis, Davis, California
95616} \affiliation{University of California, Los Angeles, Los
Angeles, California  90024} \affiliation{University of California,
San Diego, La Jolla, California  92093} \affiliation{University of
California, Santa Barbara, Santa Barbara, California 93106}
\affiliation{Instituto de Fisica de Cantabria, CSIC-University of
Cantabria, 39005 Santander, Spain} \affiliation{Carnegie Mellon
University, Pittsburgh, PA  15213} \affiliation{Enrico Fermi
Institute, University of Chicago, Chicago, Illinois 60637}
\affiliation{Comenius University, 842 48 Bratislava, Slovakia;
Institute of Experimental Physics, 040 01 Kosice, Slovakia}
\affiliation{Joint Institute for Nuclear Research, RU-141980 Dubna,
Russia} \affiliation{Duke University, Durham, North Carolina  27708}
\affiliation{Fermi National Accelerator Laboratory, Batavia,
Illinois 60510} \affiliation{University of Florida, Gainesville,
Florida  32611} \affiliation{Laboratori Nazionali di Frascati,
Istituto Nazionale di Fisica Nucleare, I-00044 Frascati, Italy}
\affiliation{University of Geneva, CH-1211 Geneva 4, Switzerland}
\affiliation{Glasgow University, Glasgow G12 8QQ, United Kingdom}
\affiliation{Harvard University, Cambridge, Massachusetts 02138}
\affiliation{Division of High Energy Physics, Department of Physics,
University of Helsinki and Helsinki Institute of Physics, FIN-00014,
Helsinki, Finland} \affiliation{University of Illinois, Urbana,
Illinois 61801} \affiliation{The Johns Hopkins University,
Baltimore, Maryland 21218} \affiliation{Institut f\"{u}r
Experimentelle Kernphysik, Universit\"{a}t Karlsruhe, 76128
Karlsruhe, Germany} \affiliation{Center for High Energy Physics:
Kyungpook National University, Daegu 702-701, Korea; Seoul National
University, Seoul 151-742, Korea; Sungkyunkwan University, Suwon
440-746, Korea; Korea Institute of Science and Technology
Information, Daejeon, 305-806, Korea; Chonnam National University,
Gwangju, 500-757, Korea} \affiliation{Ernest Orlando Lawrence
Berkeley National Laboratory, Berkeley, California 94720}
\affiliation{University of Liverpool, Liverpool L69 7ZE, United
Kingdom} \affiliation{University College London, London WC1E 6BT,
United Kingdom} \affiliation{Centro de Investigaciones Energeticas
Medioambientales y Tecnologicas, E-28040 Madrid, Spain}
\affiliation{Massachusetts Institute of Technology, Cambridge,
Massachusetts  02139} \affiliation{Institute of Particle Physics:
McGill University, Montr\'{e}al, Qu\'{e}bec, Canada H3A~2T8; Simon
Fraser University, Burnaby, British Columbia, Canada V5A~1S6;
University of Toronto, Toronto, Ontario, Canada M5S~1A7; and TRIUMF,
Vancouver, British Columbia, Canada V6T~2A3} \affiliation{University
of Michigan, Ann Arbor, Michigan 48109} \affiliation{Michigan State
University, East Lansing, Michigan  48824} \affiliation{Institution
for Theoretical and Experimental Physics, ITEP, Moscow 117259,
Russia} \affiliation{University of New Mexico, Albuquerque, New
Mexico 87131} \affiliation{Northwestern University, Evanston,
Illinois  60208} \affiliation{The Ohio State University, Columbus,
Ohio  43210} \affiliation{Okayama University, Okayama 700-8530,
Japan} \affiliation{Osaka City University, Osaka 588, Japan}
\affiliation{University of Oxford, Oxford OX1 3RH, United Kingdom}
\affiliation{Istituto Nazionale di Fisica Nucleare, Sezione di
Padova-Trento, $^y$University of Padova, I-35131 Padova, Italy}
\affiliation{LPNHE, Universite Pierre et Marie Curie/IN2P3-CNRS,
UMR7585, Paris, F-75252 France} \affiliation{University of
Pennsylvania, Philadelphia, Pennsylvania 19104}
\affiliation{Istituto Nazionale di Fisica Nucleare Pisa,
$^z$University of Pisa, $^{aa}$University of Siena and $^{bb}$Scuola
Normale Superiore, I-56127 Pisa, Italy} \affiliation{University of
Pittsburgh, Pittsburgh, Pennsylvania 15260} \affiliation{Purdue
University, West Lafayette, Indiana 47907} \affiliation{University
of Rochester, Rochester, New York 14627} \affiliation{The
Rockefeller University, New York, New York 10021}
\affiliation{Istituto Nazionale di Fisica Nucleare, Sezione di Roma
1, $^{cc}$Sapienza Universit\`{a} di Roma, I-00185 Roma, Italy}

\affiliation{Rutgers University, Piscataway, New Jersey 08855}
\affiliation{Texas A\&M University, College Station, Texas 77843}
\affiliation{Istituto Nazionale di Fisica Nucleare Trieste/Udine,
I-34100 Trieste, $^{dd}$University of Trieste/Udine, I-33100 Udine,
Italy} \affiliation{University of Tsukuba, Tsukuba, Ibaraki 305,
Japan} \affiliation{Tufts University, Medford, Massachusetts 02155}
\affiliation{Waseda University, Tokyo 169, Japan} \affiliation{Wayne
State University, Detroit, Michigan  48201} \affiliation{University
of Wisconsin, Madison, Wisconsin 53706} \affiliation{Yale
University, New Haven, Connecticut 06520}
\author{T.~Aaltonen}
\affiliation{Division of High Energy Physics, Department of Physics,
University of Helsinki and Helsinki Institute of Physics, FIN-00014,
Helsinki, Finland}
\author{J.~Adelman}
\affiliation{Enrico Fermi Institute, University of Chicago, Chicago,
Illinois 60637}
\author{T.~Akimoto}
\affiliation{University of Tsukuba, Tsukuba, Ibaraki 305, Japan}
\author{B.~\'{A}lvarez~Gonz\'{a}lez$^s$}
\affiliation{Instituto de Fisica de Cantabria, CSIC-University of
Cantabria, 39005 Santander, Spain}
\author{S.~Amerio$^y$}
\affiliation{Istituto Nazionale di Fisica Nucleare, Sezione di
Padova-Trento, $^y$University of Padova, I-35131 Padova, Italy}

\author{D.~Amidei}
\affiliation{University of Michigan, Ann Arbor, Michigan 48109}
\author{A.~Anastassov}
\affiliation{Northwestern University, Evanston, Illinois  60208}
\author{A.~Annovi}
\affiliation{Laboratori Nazionali di Frascati, Istituto Nazionale di
Fisica Nucleare, I-00044 Frascati, Italy}
\author{J.~Antos}
\affiliation{Comenius University, 842 48 Bratislava, Slovakia;
Institute of Experimental Physics, 040 01 Kosice, Slovakia}
\author{G.~Apollinari}
\affiliation{Fermi National Accelerator Laboratory, Batavia,
Illinois 60510}
% ...

\collaboration{CDF Collaboration\footnote{With visitors from
$^a$University of Massachusetts Amherst, Amherst, Massachusetts
01003, $^b$Universiteit Antwerpen, B-2610 Antwerp, Belgium,
$^c$University of Bristol, Bristol BS8 1TL, United Kingdom,
$^d$Chinese Academy of Sciences, Beijing 100864, China, $^e$Istituto
Nazionale di Fisica Nucleare, Sezione di Cagliari, 09042 Monserrato
(Cagliari), Italy, $^f$University of California Irvine, Irvine, CA
92697, $^g$University of California Santa Cruz, Santa Cruz, CA
95064, $^h$Cornell University, Ithaca, NY  14853, $^i$University of
Cyprus, Nicosia CY-1678, Cyprus, $^j$University College Dublin,
Dublin 4, Ireland, $^k$University of Edinburgh, Edinburgh EH9 3JZ,
United Kingdom, $^l$University of Fukui, Fukui City, Fukui
Prefecture, Japan 910-0017 $^m$Kinki University, Higashi-Osaka City,
Japan 577-8502 $^n$Universidad Iberoamericana, Mexico D.F., Mexico,
$^o$Queen Mary, University of London, London, E1 4NS, England,
$^p$University of Manchester, Manchester M13 9PL, England,
$^q$Nagasaki Institute of Applied Science, Nagasaki, Japan,
$^r$University of Notre Dame, Notre Dame, IN 46556, $^s$University
de Oviedo, E-33007 Oviedo, Spain, $^t$Texas Tech University,
Lubbock, TX  79609, $^u$IFIC(CSIC-Universitat de Valencia), 46071
Valencia, Spain, $^v$University of Virginia, Charlottesville, VA
22904, $^w$Bergische Universit\"at Wuppertal, 42097 Wuppertal,
Germany, $^{ee}$On leave from J.~Stefan Institute, Ljubljana,
Slovenia, }} \noaffiliation

\date{\today}

\begin{abstract}

We present a measurement of the $\ttbar$ differential cross section
with respect to the $\ttbar$ invariant mass, $\dxs$, in $\ppbar$
collisions at $\sqrt{s}=1.96$ TeV using an integrated luminosity of
$2.7~\invfb$ collected by the CDF II experiment. The $\ttbar$
invariant mass spectrum is sensitive to a variety of exotic
particles decaying into $\ttbar$ pairs. The result is consistent
with the standard model expectation, as modeled by \texttt{PYTHIA}
with \texttt{CTEQ5L} parton distribution functions.

\end{abstract}

\maketitle

\newpage

 The top quark is the only known fermion with a
mass near the electroweak symmetry breaking (EWSB) scale~\cite{PDG}.
As such, it plays a special role in many beyond the standard model
(BSM) theories of EWSB.  In the standard model (SM) the Higgs boson
is responsible for EWSB and the generation of the fermion masses,
but it has not yet been observed.  In models with top condensation,
such as technicolor and topcolor, the role of the SM Higgs boson is
filled by a composite particle that is a bound state of top
quarks~\cite{TopCond}. These models predict additional heavy gauge
bosons that couple strongly to top quarks.  The hierarchy problem,
also unresolved in the SM, has recently been addressed by models
with extra dimensions, such as the Randall-Sundrum
(RS)~\cite{RandallSundrum} and ADD models~\cite{ADD}. In these
models TeV-scale gravitons can decay, in some cases preferentially,
to $\ttbar$ pairs~\cite{Fitz}. In all of these cases the production
of $\ttbar$ pairs at hadron colliders through BSM mechanisms
distorts the $\ttbar$ invariant mass spectrum relative to the SM
expectation, as recently reviewed in~\cite{Fabio}.

In this Letter we report on the first measurement of the $\ttbar$
differential cross section with respect to the $\ttbar$ invariant
mass, $\dxs$. The analysis uses an integrated luminosity of
2.70$\pm$0.16~fb$^{-1}$~\cite{klimenko} collected with the CDF II
detector between March 2002 and April 2008.  Full details of the
analysis presented here are given in~\cite{AliceThesis}. Previous
published studies have focused on searches for resonances in the
$\Mtt$ spectrum~\cite{ZprimePRD}, and placed a lower limit of
720~$\GeVcc$ on the mass of a putative $Z^\prime$ boson decaying
preferentially to $\ttbar$.  In this Letter we take a different
approach in which we test the $\Mtt$ spectrum, generically, for
consistency with the SM. In this way we are potentially sensitive to
broad enhancements of the spectrum and interference
effects~\cite{Fabio}, as well as to narrow resonances.

The CDF II detector is described in detail elsewhere~\cite{CDF}. The
components relevant to this analysis include the silicon vertex
detector (SVX), the central outer tracker (COT), the central
electromagnetic and hadronic calorimeters, the central muon
detectors, and the luminosity counters.

Because it allows reconstruction of the final state with good
resolution, and because of the good signal to background ratio, we
use the ``lepton+jets" decay mode of the $\ttbar$ pair in this
study. It consists of four energetic jets, two of which originate
from bottom quarks and two from the hadronic $W$-boson decay, a
charged lepton with large transverse momentum ($\Pt$), and a large
transverse momentum imbalance ($\met$) from the undetected neutrino
from the leptonic $W$-boson decay~\cite{MET}. Extra jets may appear
from initial- or final-state radiation (ISR or FSR). A $\ttbar$
event may also be observed with fewer than four jets if a jet is not
reconstructed or is merged with another jet in the event. Monte
Carlo (MC) simulations of $\ttbar$ production are generated using
the {\tt PYTHIA} MC program~\cite{Pythia} with the {\tt
CTEQ5L}~\cite{CTEQ} parton distribution functions (PDFs).  The
decays of heavy quarks ($b$ and $c$) are modeled using {\tt
EVTGEN}~\cite{EvtGen}. The {\tt HERWIG} MC program~\cite{Herwig} is
used for  studies of the systematic effects of the hadronization
model.  The $\ttbar$ MC samples are generated with a top quark mass
of 175 GeV/c$^2$. The results presented here are insensitive to
changes of the generated top quark mass of a few GeV/c$^2$.

Events from $\ppbar$ collisions are selected with an inclusive
lepton trigger that requires an electron (muon) candidate with $\Et
>$ 18 GeV ($\Pt >$ 18 $\GeVc$).  From the triggered events the
signal sample is selected offline by requiring an isolated electron
(muon) with $\Et >$ 20 GeV ($\Pt >$ 20 $\GeVc$).  The isolation
criterion requires $I <$ 0.1, where $I$ is defined as the
calorimeter transverse energy in a cone of opening radius $\Delta
R\equiv\sqrt{(\Delta\eta)^2 + (\Delta\phi)^2}=0.4$ around the lepton
direction (exclusive of the lepton energy), divided by the electron
(muon) $\Et$ ($\Pt$).  We further require $\met >$ 20 GeV and at
least 4 jets each with $\Et~>~$20 GeV and $\mid\!\eta\!\mid <2.0$.
Jets are identified using a fixed-cone algorithm with a cone size of
$\Delta R=0.4$ and are constrained to originate at the $\ppbar$
collision vertex. Their energies are corrected to account for
detector response variations in $\eta$, calorimeter gain
instability, nonlinearity of calorimeter energy response, multiple
interactions in an event, and for energy loss in un-instrumented
regions of the detector. For events with more than four jets with
$\Et~>~$20 GeV, we use the four highest-$\Et$ jets in the $\Mtt$
reconstruction.  Missing transverse energy is corrected to account
for the shifts in jet energies due to the jet energy corrections
described above. $Z$-boson candidate events are rejected by removing
events containing a second isolated high-$\Pt$ lepton. To suppress
the background from direct production of a $W$ boson and multiple
jets ($W$+jets events) we require that at least one jet in the event
have an identified displaced secondary vertex, consistent with the
decay of a $B$ hadron. We label such jets, and the events that
contain them, as ``$b$-tagged". The events selected prior to the
$b$-tag requirement are called ``pretag" events. We observe 2069
pretag, and 650 $b$-tagged, events.

 The $\ttbar$ signature described above can be mimicked by
several processes, including diboson ($WW$, $WZ$, $ZZ$), single-top,
$Z$+jets, and $W$+jets production, as well as processes without
vector bosons to which we refer, generically, as ``QCD" backgrounds.
The diboson and single-top quark yields are predicted using {\tt
PYTHIA} and {\tt MADEVENT}~\cite{MadEvent} MC samples, respectively,
each normalized to the theoretical cross
sections~\cite{Campbell,Sullivan}. The residual $Z$+jets background
is modeled using {\tt ALPGEN}~\cite{Alpgen}, with the parton
showering and underlying event model from {\tt PYTHIA}. The QCD
background typically has lower $\met$ than events with real $W$
bosons and is evaluated by fitting the $\met$ distribution using
templates for QCD and $W$+jets sources and extrapolating the QCD
fraction into the high-$\met$ signal region. {\tt ALPGEN} is also
used in the evaluation of the dominant background from $W$+jets
production. The $W$+jets background is determined separately for
events with and without heavy-flavor jets. For events with
heavy-flavor jets we use the {\tt ALPGEN} simulation to determine
the fraction, in each jet multiplicity bin, of $W$+jets events that
are $W\bbbar$, $W\ccbar$ or $Wc$.  This fraction is then increased
by a correction factor, determined by comparing measured and {\tt
ALPGEN}-predicted heavy-flavor (HF) fractions in $W$+1 jet data. The
number of pretag $W+$heavy-flavor events is normalized to the total
number of $W$+jets events in each jet multiplicity bin of the data
using the modified {\tt ALPGEN} fractions.  The background
contribution from these events is given by the pretag number of
events times a MC-derived tagging efficiency.  Events without
heavy-flavor jets can enter the signal sample if one of the jets is
mistakenly $b$-tagged. Such events are called ``mistags", and they
occur primarily due to tracking errors, with a smaller contribution
from interactions in the material of the detector, and $K_S$ and
$\Lambda$ decays. The background due to mistags in $W$+jets events
is evaluated using a measurement of the rate of mistags derived from
multijet data~\cite{secvtx}. The mistag rate is then applied to the
number of pretag $W$+jets events, with no heavy-flavor jets, in the
data. This pretag number is calculated, in each jet multiplicity
bin, from the total number of candidate events corrected for the
contributions from non-$W$+jets and from $W$+heavy-flavor jets. The
observed event yields and background predictions are given in
Tab.~\ref{tab:obs}, where the line labeled `Other' includes
dibosons, $Z$+jets and single top.

\begin{table}
\begin{tabular}{lcc}
\hline \hline
Process & 4 jets & $\geq5$ jets \\
\hline
$W$+HF & 58.0 $\pm$ 12.2 & 11.6 $\pm$ 2.9 \\
Mistags & 18.9 $\pm$ 4.8 & 3.5 $\pm$ 1.6 \\
QCD & 20.9 $\pm$ 17.5 & 6.4 $\pm$ 6.0 \\
Other & 13.9 $\pm$ 0.8 & 3.1 $\pm$ 0.2 \\
$t\bar{t}$ (6.7pb) & 358.6 $\pm$ 49.7 & 121.5 $\pm$ 16.8 \\
\hline
Total Prediction & 470 $\pm$ 57 & 146 $\pm$ 19 \\
Observed & 494 & 156 \\
\hline
\hline
\end{tabular}
\caption{Summary of sample composition~\cite{3fbsecvtx}.}
\label{tab:obs}
\end{table}

The precision of the measurement of $\Mtt$ depends on the
understanding of the jet energy scale (JES).  To reduce the
uncertainty on the JES, we adopt an approach first used
in~\cite{TMTPRL} and use the measured invariant mass of the
hadronically-decaying $W$ boson to constrain the JES. For events
with two $b$-tagged jets, the two un-tagged jets are chosen as the
jets from the $W$ boson decay.  For events with a single $b$-tagged
jet, the pair with invariant mass closest to the expected mean value
from $W$ boson decays is chosen. There are 503 single-tagged and 147
double-tagged events in the sample. An unbinned maximum likelihood
fit, using MC templates for the dijet invariant mass distribution,
for both signal and background, returns the best-fit JES and its
uncertainty.  The fit value of the JES is subsequently used in the
analysis.  The uncertainty returned by this procedure is
approximately a factor of two lower than the nominal uncertainty on
the JES.

We reconstruct $\Mtt$, the $\ttbar$ invariant mass, using the
four-vectors of the $b$-tagged jet and the three remaining leading
jets in the event, the lepton and the transverse components of the
neutrino momentum, given by $\met$. We divide the $\Mtt$
distribution into nine bins between 0 and 1400~$\GeVcc$, with bin
widths ranging from 50~$\GeVcc$ for bins for which a large number of
events are expected to 600~$\GeVcc$ for the highest bin. The
resolution in $\Mtt$ is somewhat smaller than the bin width, ranging
from 11\% near the peak to 15\% at high mass. We subtract from the
bulk $\Mtt$ distribution the expected contribution from the
backgrounds listed in Tab.~\ref{tab:obs}, which is modeled using the
Monte Carlo samples described above.  The resulting $\Mtt$ signal
distribution suffers from resolution smearing and is corrected using
a regularized unfolding technique, described below, which also
accounts for the longitudinal component of the neutrino momentum.

In order to extract the true underlying $\Mtt$ distribution from the
background-subtracted reconstructed distribution, we use the MC to
create a response matrix $\ahat$, such that $\ahat x = d$ where $x$
is the true, binned distribution and $d$ is the measured, binned
distribution.  Due to statistical fluctuations in bins with small
numbers of events, inverting the response matrix $\ahat$ to solve
for $x$ given $d$ often yields spurious results.  Instead we use
singular-value decomposition (SVD) unfolding, as described
in~\cite{SVD}, where the solution is regularized by populating the
response matrix with event multiplicities instead of probabilities.
The application to this analysis is described in detail
in~\cite{AliceThesis}.

From the unfolded $\Mtt$ distribution, we calculate the differential
cross section according to
\begin{equation}\label{eq:diffxs}\nonumber
\frac{d\sigma^{i}}{d\Mtt} =
\frac{N_{i}}{\acc\int\mathcal{L}dt\cdot\Delta_{\Mtt}^{i}},
\end{equation}
where $N_{i}$ is the background-subtracted, unfolded, number of
events observed in each bin;  $\acc$ is the acceptance in bin $i$;
$\Delta_{\Mtt}^{i}$ is the width of bin $i$; and
$\int\mathcal{L}{\rm dt}$ is the integrated luminosity.  The
acceptance is measured from a mixture of data and MC.  We use {\tt
PYTHIA} with a {\tt GEANT}-based~\cite{secvtx} detector simulation
to measure the geometric and kinematic acceptance. The lepton
trigger and identification efficiencies are measured in data using
$Z\rightarrow\ell\ell$ decays.  We account for the difference in
efficiency for identifying an isolated high-$\Pt$ lepton in data and
MC with a scale factor. Similarly we use a scale factor to correct
for the difference in efficiency in data and MC for tagging a
$b$-jet. The efficiency in data is determined in a
heavy-flavor-enriched data sample of low-$\Pt$ electrons, from the
semi-leptonic decay of $B$ hadrons.

Our systematic uncertainties arise from MC modeling of the
acceptance, true and reconstructed $\Mtt$ distributions, and
background distributions. In addition, the uncertainties of our
efficiency of lepton identification, $b$-tagging efficiency, and
integrated luminosity affect the measurement.  The lepton
identification uncertainty arises due to the extrapolation from
$Z\rightarrow\ell\ell$ events, where the efficiency is measured in
data, to the higher multiplicity $\ttbar$ environment. The
uncertainty on the $b$-tagging efficiency is largely due to the
limited number of events in the data sample that is used. These
uncertainties, together with a small uncertainty due to the finite
size of the MC simulation used to calculate the acceptance, comprise
the acceptance uncertainty in Tab.~\ref{tab:Syst}.

We consider several variations to the MC model of the signal and
background.  For the signal MC simulation we compare the results
using {\tt HERWIG} to the default {\tt PYTHIA} generator.  The
uncertainty due to the limited knowledge of ISR is constrained by
studies of radiation in Drell-Yan events in the data.  We vary both
ISR and FSR within these limits and add the deviations from the
nominal value in quadrature. The uncertainty due to possible
differences in the PDFs from the nominal {\tt CTEQ5L} PDF is
evaluated by varying the PDF using the 20 {\tt CTEQ5L} eigenvectors,
which represent 90\% C.L. variations. The deviations from the
nominal values are added in quadrature with deviations measured
using the {\tt MRST} PDF~\cite{MRST} with two alternate choices for
the strong coupling constant. The uncertainty on the background
prediction consists of two pieces: the uncertainty on the background
normalization, given in Tab.~\ref{tab:obs}, and a background shape
systematic for the MC modeling of the shapes.  The systematic
uncertainty due to the JES includes a generic energy-correction
systematic uncertainty as well as a contribution from the modeling
of the $b$-jet energy scale.  The effects of each systematic
uncertainty on the measurement are evaluated using a
pseudo-experiment approach. Pseudo-experiments are performed for
each variation described above and the difference between the mean
$\dxs$ in each bin with the shifted parameters and the default model
is taken as the systematic uncertainty in that bin. The results are
presented in Tab.~\ref{tab:Syst}. The dominant systematic
uncertainty is the uncertainty on the PDF set. This is expected as
the tail of the $\Mtt$ spectrum is very sensitive to the PDFs.  The
$6\%$ uncertainty on the luminosity measurement in each
bin~\cite{klimenko} is not included in the total in
Tab.~\ref{tab:Syst}.  Two effects cause the uncertainty in the bins
between 400$~\GeVcc$ and 550$~\GeVcc$ to be somewhat smaller than
outside of that range.  One is the turn-on threshold of the $\Mtt$
spectrum, which is insensitive to systematic variations because we
fix the top quark mass at 175$~\GeVcc$.  The second is the PDF
uncertainty, which is much greater at large $\Mtt$ than at small
$\Mtt$.
\begin{table*}[htbp]
\begin{center}
\begin{tabular}{lrrrrrrrrr}
\hline \hline $\Mtt$ [$\GeVcc$] & 0-350 & 350-400 & 400-450 &
450-500
& 500-550 & 550-600 & 600-700 & 700-800& 800-1400\\ \hline
MC Gen.&0.7 &2.4 &5.3 &5.7 &4.6 &3.3 &1.4 &1.0 &1.0\\
ISR/FSR&1.5 &1.3 &0.8 &0.2 &1.1 &2.1 &2.0 &2.2 &3.3\\
JES & 8.2 & 6.3 & 4.1 & 3.1 & 1.7 & 2.3 & 4.6 & 7.5 & 9.1 \\
Backgrounds & 10.3 & 7.4 & 2.4 & 1.7 & 3.0 & 4.0 & 4.5 & 5.1 & 5.4 \\
Acceptance &4.5 &4.4 &4.3 &4.4 &4.6 &4.6 &4.4 &4.0 &3.8\\
PDF Set &7.7 &6.1 &3.0 &1.0 &4.8 &9.3 &14.0 &17.4 &18.8\\ \hline
Total &16.0 &12.6 &8.9 &8.1 &8.9 &12.0 &16.1 &20.1 &22.2\\
\hline \hline\end{tabular}
\end{center}
\caption{Summary of systematic uncertainties (in \%) in each bin.
The $6\%$ uncertainty on the integrated luminosity is not included
in the total.} \label{tab:Syst}\rm\end{table*}

\begin{figure}
\begin{center}
\includegraphics[width=3.375in]{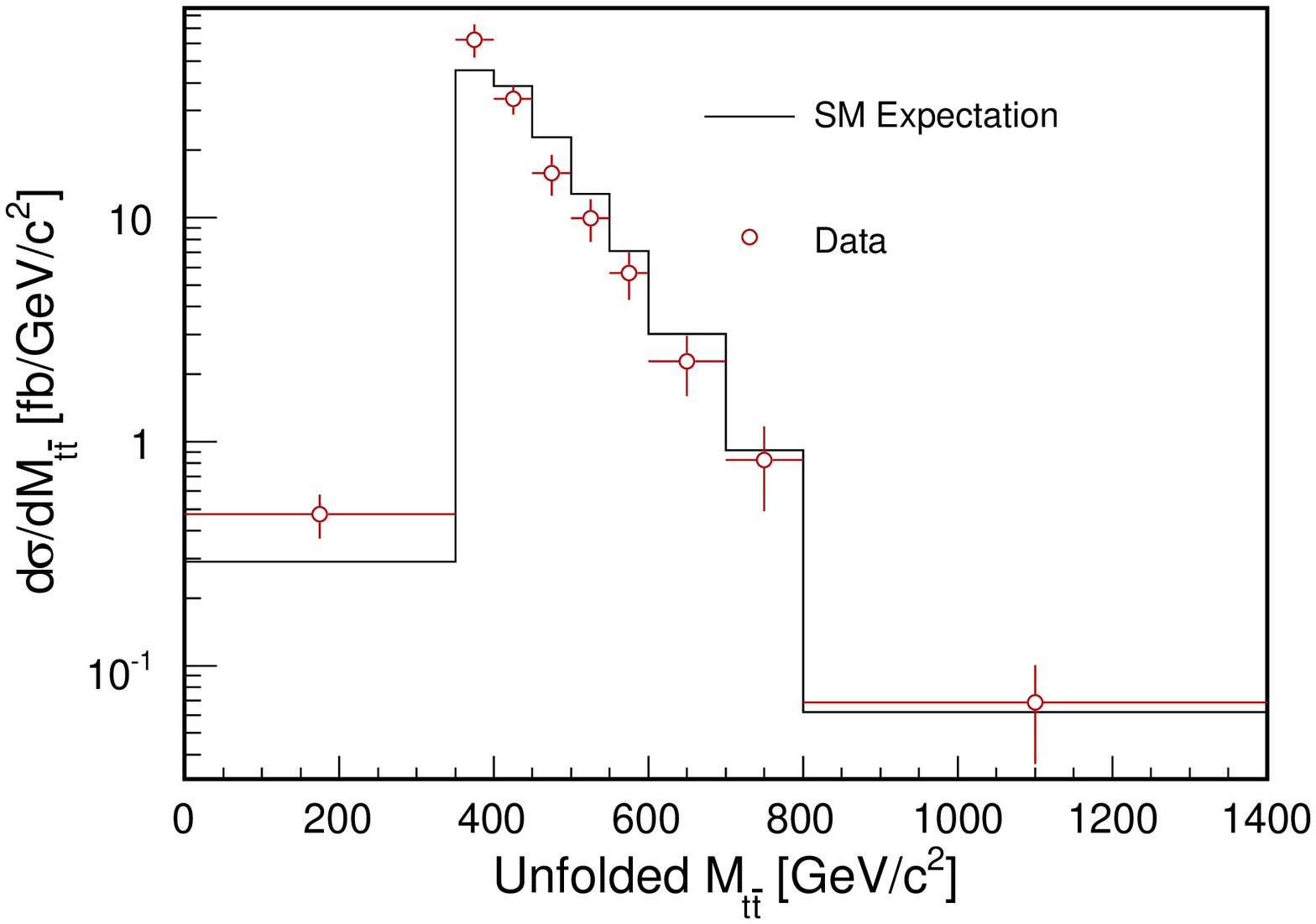}
\caption{$\dxs$ measured with 2.7~$\invfb$ of integrated luminosity.
}\label{fig:dxs}
\end{center}
\end{figure}

The measured $\dxs$ is shown in Fig.~\ref{fig:dxs} and tabulated in
Tab.~\ref{tab:dxs}.
\begin{table}[htbp]
\begin{center}
\begin{tabular}{lcl}
\hline \hline $\Mtt$ [$\GeVcc$]& $\acc$ & $\dxs$ [$\fbpergev$]\\
\hline
$\leq$ 350& 0.016$\pm$0.001~~~& 0.47 $\pm$ 0.07 $\pm$ 0.08 $\pm$ 0.03 \\
350-400& 0.023$\pm$0.001~~~& 62.3 $\pm$ 7.0  $\pm$ 7.9$\pm$ 3.7\\
400-450& 0.026$\pm$0.001~~~& 33.8 $\pm$ 4.0  $\pm$ 3.0  $\pm$ 2.0 \\
450-500& 0.027$\pm$0.001~~~& 15.8 $\pm$ 3.0  $\pm$ 1.3  $\pm$ 0.9 \\
500-550& 0.029$\pm$0.001~~~& 9.9 $\pm$ 2.0  $\pm$ 0.9  $\pm$ 0.6\\
550-600& 0.030$\pm$0.001~~~& 5.7 $\pm$ 1.2  $\pm$ 0.7  $\pm$ 0.3 \\
600-700& 0.030$\pm$0.001~~~& 2.3 $\pm$ 0.6  $\pm$ 0.4  $\pm$ 0.1 \\
700-800& 0.030$\pm$0.001~~~& 0.8 $\pm$ 0.3  $\pm$ 0.2  $\pm$ 0.1 \\
800-1400&\hspace{-.14in} 0.023$\pm$0.001& 0.068 $\pm$ 0.032 $\pm$ 0.015 $\pm$ 0.004 \\
\hline \hline \multicolumn{3}{c}{Integrated Cross Section [pb]
$6.9\pm1.0$ (stat.+JES)}\\ \hline \hline
\end{tabular}
\end{center}\caption{The acceptance and measured differential cross section in each
bin. The uncertainties on the cross-section values are,
respectively, statistical+JES, systematic and
luminosity.}\label{tab:dxs}\rm\end{table}

We check consistency with the SM prediction using the
Anderson-Darling (AD) statistic~\cite{ADStat}, which places an
emphasis on potential discrepancies in the tail of the $\Mtt$
distribution. The distribution of the AD statistic for this analysis
is rapidly falling, with small values corresponding to more likely
results. Using MC simulations, we optimize the bin range of the
Anderson-Darling statistic for maximum sensitivity to new physics
and a minimum of false positives and find $\Mtt\geq450~\gev$ to be
the most sensitive region of $\Mtt$.  We perform pseudo-experiments
using the SM MC distributions of $\Mtt$ with the sample composition
given in Tab.~\ref{tab:obs}. We calculate a p-value by taking the
fraction of pseudo-experiments with a larger observed ({\it i.e.}
less likely in the SM) Anderson-Darling statistic than that in data.
The observed p-value is 0.28.  We conclude that there is no evidence
of non-SM physics in the $\Mtt$ distribution.

We thank Rikkert Frederix, Fabio Maltoni, and Tim Stelzer for
stimulating discussions, advice, and help with MadEvent generation.
We thank the Fermilab staff and the technical staffs of the
participating institutions for their vital contributions. This work
was supported by the U.S. Department of Energy and National Science
Foundation; the Italian Istituto Nazionale di Fisica Nucleare; the
Ministry of Education, Culture, Sports, Science and Technology of
Japan; the Natural Sciences and Engineering Research Council of
Canada; the National Science Council of the Republic of China; the
Swiss National Science Foundation; the A.P. Sloan Foundation; the
Bundesministerium f\"ur Bildung und Forschung, Germany; the Korean
Science and Engineering Foundation and the Korean Research
Foundation; the Science and Technology Facilities Council and the
Royal Society, UK; the Institut National de Physique Nucleaire et
Physique des Particules/CNRS; the Russian Foundation for Basic
Research; the Ministerio de Ciencia e Innovaci\'{o}n, and Programa
Consolider-Ingenio 2010, Spain; the Slovak R\&D Agency; and the
Academy of Finland.


\begin{thebibliography}{99}
\bibitem{PDG} C. Amsler {\it et al.}, Phys. Lett. B {\bf 667}, 1 (2008).

\bibitem{TopCond} G. Cvetic, Reviews of Modern Physics {\bf 71}, 513(1999).

\bibitem{RandallSundrum} L. Randall and R. Sundrum, Phys. Rev. Lett., {\bf 83}, 3370 (1999).

\bibitem{ADD} N. Arkani-Hamed, S. Dimopoulos and G. R. Dvali, Phys. Lett. B {\bf 429}, 263
(1998).

\bibitem{Fitz} L. Fitzpatrick, J. Kaplan, L. Randall, L. Wang, JHEP {\bf 0709}, 013
(2007).

\bibitem{Fabio} R. Frederix and F. Maltoni, JHEP {\bf 0901}, 047 (2009).

    \bibitem{klimenko} S. Klimenko, J. Konigsberg, and T.M. Liss,
    Fermilab-FN-0741.

    \bibitem{AliceThesis} Alice Patricia Bridgeman, Measurement of the
    $\ttbar$ Differential Cross Section, $\lowercase{d\sigma/d}M_{\lowercase{{t\overline{t}}}}$,
     in $\ppbar$ Collisions at
    $\sqrt{s}=1.96$ TeV, Ph.D Thesis,
    University of Illinois at Urbana-Champaign, 2008.  FERMILAB-THESIS-2008-50

    \bibitem{ZprimePRD} T. Aaltonen {\it et al.}, (CDF Collaboration), Phys. Rev. D {\bf 77}, 051102(R)
    (2008); V.M. Abazov {\it et al.}, (D\O~Collaboration), Phys. Lett. B {\bf 668}, 98
    (2008).



    \bibitem{CDF}
    The CDF II Detector Technical Design Report,
    Fermilab-Pub-96/390-E; D. Amidei {\it et al.},
Nucl. Instum. Methods Phys. Res. A {\bf 350}, 73 (1994); F. Abe {\it
et al.}, Phys. Rev. D {\bf 52}, 4784 (1995); P. Azzi {\it et al.},
Nucl. Instrum. Methods Phys. Res. A {\bf 360}, 137 (1995); D. Acosta
{\it et al.}, Phys. Rev. D {\bf 71}, 032001 (2005).

\bibitem{MET} We use a coordinate system where the $z$-axis
    is in the direction of the proton beam, and $\phi$ and $\theta$ are the azimuthal
    and polar angles respectively.  The pseudorapidity, $\eta$, is defined as
    $\eta =-\ln(\tan\frac{\theta}{2})$.  The transverse momentum of a charged particle is
    $\Pt = P\sin\theta$, where $P$ represents the measured momentum of the charged-particle
    track.  The analogous quantity using calorimeter energies, defined as $E_T =
    E\sin\theta$,
    is called transverse energy. The missing transverse energy is defined as
    $\met = -\mid\sum_i E_T^i\hat{n}_i\mid$ where $E_T^i$ is the
    magnitude of the transverse energy contained in each calorimeter
    tower $i$ in the pseudorapidity region $\mid\eta\mid <3.6$, and
    $\hat{n}_i$ is the direction unit vector of the tower in the
    plane transverse to the beam direction.

\bibitem{Pythia}T. Sj\"{o}strand, S. Mrenna, and P. Skands, JHEP {\bf 0605}, 026
(2006).  We use {\sc PYTHIA} version 6.216.

\bibitem{CTEQ}H.~L.~Lai {\it et al.}  (CTEQ Collaboration),
  Eur.\ Phys.\ J.\  C {\bf 12}, 375 (2000).

  \bibitem{EvtGen}D. J. Lange, Nucl. Instrum. Meth. A {\bf 462}, 152 (2001).

\bibitem{Herwig}G. Corcella {\it et al.}, JHEP {\bf 0101}, 010
(2001).

\bibitem{MadEvent}J. Alwall {\it et al.},
JHEP {\bf 0709}, 028 (2007).

\bibitem{Campbell} J. M. Campbell and R. K. Ellis,  Phys. Rev. D
{\bf 60}, 113006 (1999).

\bibitem{Sullivan} Z. Sullivan,  Phys. Rev. D {\bf 70}, 114012 (2004).

\bibitem{Alpgen}M.L. Mangano {\it et al.}, JHEP
{\bf 0307}, 001 (2003).

\bibitem{secvtx} D. Acosta {\it et al.}, (CDF Collaboration), Phys. Rev. D {\bf 71}, 052003 (2005).

\bibitem{3fbsecvtx} CDF Conference Note 9462,\\
    \verb"http://www-cdf.fnal.gov/physics/new/top/2008/"\\
    \verb"xsection/ttbar_secvtx_3invfb/"

    \bibitem{TMTPRL} A. Abulencia {\it et al.}, (CDF Collaboration), Phys. Rev. Lett.
    {\bf 96}, 022004 (2006); A. Abulencia {\it et al.}, (CDF Collaboration), Phys. Rev. D
    {\bf 73}, 032003 (2006).

    \bibitem{SVD} A. Hocker and V. Kartvelishivili, hep-ph/9509307 (1995).

\bibitem{MRST} A. D. Martin {\it et al.}, Eur. Phys. J. C {\bf 14}, 133
(2000).

\bibitem{ADStat}
    T. W. Anderson and D. A. Darling, The Annals of Math.
    Stat. {\bf 23}, 193 (1952).

\end{thebibliography}
\end{document}